\documentclass{emulateapj}

\newcommand{\mr}{\mathrm}

\shorttitle{{\it RHESSI} quiet Sun observations.}

\shortauthors{Hannah et al.}

\begin{document}

\title{Constraining the hard X-ray properties \\of the quiet Sun with new RHESSI
observations}
\author{I. G. Hannah} \affil{School of Physics \& Astronomy, \\University of
Glasgow, Glasgow G12 8QQ, UK}\email{iain@astro.gla.ac.uk}
\author{H. S. Hudson, G. J. Hurford}
\affil{Space Sciences Laboratory, \\University of California, Berkeley, CA,
94720-7450, USA} \email{\\hudson@ssl.berkeley.edu, hurford@ssl.berkeley.edu}
\author{R. P. Lin}\affil{Physics Department \& Space Sciences Laboratory,
\\University of California, Berkeley, CA, 94720-7450, USA;
\\School of Space Research, Kyung Hee University,
Korea}\email{rplin@ssl.berkeley.edu }

\begin{abstract}
We present new RHESSI upper limits in the 3-200 keV energy range for solar
hard X-ray emission in the absence of flares and active regions, i.e. the quiet
Sun, using data obtained between July 2005 and April 2009. These new limits,
substantially deeper than any previous ones, constrain several physical
processes that could produce hard X-ray emission. These include cosmic-ray
effects and the generation of axions within the solar core. The data also limit
the properties of ``nanoflares'', a leading candidate to explain coronal heating.
We find it unlikely for nanoflares involving nonthermal effects to heat the
corona because such events would require a steep electron spectrum
E$^{-\delta}$ with index $\delta
> 5$ extending to very low energies ($<1 $~keV), into the thermal energy range.
We also use the limits to constrain the parameter space of an isothermal
model and coronal thin-target emission models (powerlaw and kappa
distributions).\end{abstract}


\section{Introduction}

To a hard X-ray telescope much more sensitive than RHESSI \citep{lin2002}, the
quiet Sun, i.e. free of flares and active regions, should appear dark against the
diffuse cosmic X-ray sky. But how faint can the solar atmosphere itself be?
Intense magnetic fields collect in the network of convective motions at the
photosphere, and a wide variety of transient phenomena occur all the time even
in the absence of sunspots or other major kinds of solar activity. The high
temperature (several MK) of the corona itself has always posed a problem, with
abundant literature devoted to finding the source of energy involved in
maintaining it. An often cited idea is that of a large number of events too weak
to detect individually, but pervading the volume of the corona and extracting the
energy of its magnetic field bit by bit -- the ``nanoflares'' discussed by
\cite{parker1988}. A nanoflare population may operate in a similar manner to
the suggested active-region nanoflares \citep[e.g.][]{1997ApJ...478..799C} or be
considerably smaller versions of traditional active-region flares
\citep[e.g.][]{2008ApJ...677..704H} but they would need to exist in the absence
of active regions given that the corona remains consistently hot during quiet
periods.

Such short-duration transient heating events, occurring on an Alfv{\' e}n time
scale, would temporarily produce a higher temperature than the mean (greater
than a few MK) and the resulting differential emission measure DEM of an
ensemble of such events in the steady state must therefore extend to higher
temperatures \cite[e.g.][]{1994ApJ...422..381C}, producing soft X-rays (SXR),
emission typically below a few to 10 keV. Or if they operated in a
similar manner to active-region flares, where accelerated electrons heat the
chromospheric material, then they would produce a faint hard X-ray (HXR)
signature via nonthermal bremsstrahlung, emission typically above a few
to 10 keV. Either way a quiet-Sun nanoflare population would likely produce SXR
and HXR emission above 3~keV, an energy range observable by RHESSI.

Other X-ray observations of the quiet Sun have either provided isothermal model
fits to the limiting SXR emission
\citep{2000ApJ...528..537P,pevtsov2001,2010EOSTr..91...73S} or upper limits to
the HXR emission \citep{peterson1966,feffer1997}. RHESSI uniquely bridges the
SXR to HXR energy range and so is an ideal tool to investigate solar thermal and
nonthermal emission. However its imaging is optimised for flare observations
and so in its normal mode of operation is ill-suited to observing the weak,
spatially wide-spread signal from the quiet Sun. Instead an off-pointing mode of
operation was developed termed \emph{fan-beam modulation}
\citep{hannah_fbm},  (see further details in \S\ref{sec:method}) which allows a
weak full-disc signal to be investigated. This produced more stringent upper
limits to the quiet Sun X-ray emission between 3-200 keV
\citep{2007ApJ...659L..77H}, covering a wider energy range than previously
found \citep{peterson1966,feffer1997}.

In this paper we present a two-fold improvement over this analysis. First we
present deeper RHESSI quiet-Sun upper limits found using offpointing data from
the whole of the exceptional minimum of Solar Cycle 23, 2005 to 2009 (the
previous analysis covered only 2005 to 2006). Secondly we use these limits to
investigate the thermal (\S\ref{sec:thermal}) and nonthermal
(\S\ref{sec:nonthermal}) properties of a possible nanoflare population. In the
latter case we investigate whether they can satisfy the coronal heating
requirement \citep{1977ARA&A..15..363W}. We also
consider, in \S\ref{sec:thin}, the upper limits in the terms of possible
coronal thin-target emission.

Outside the domain of solar activity, there are other mechanisms that would
produce HXR emission. At some level the high-energy galactic cosmic rays will
result in X-ray emissions from the photosphere
\citep[e.g.][]{seckel1991,2007A&A...462..763M}. The $\gamma$-ray emission
from cosmic rays interacting with the solar atmosphere have recently been
observed with FERMI \citep{2009arXiv0912.3775O}. The cosmic X-ray
background, known to be of extragalactic origin, is bright and has a relatively
flat (hard) spectrum. It should be blocked by the solar disk, yet produce a diffuse
component via Compton scattering \citep[e.g.][]{2008MNRAS.385..719C}. A
well-defined X-ray source could also result from axion production in the core of
the Sun, converting via interactions with the magnetic field in the solar
atmosphere \citep{sikivie1983,carlson1996}. We discuss briefly in
\S\ref{sec:other} the interpretation of the RHESSI limits in terms of these other
emission mechanisms.

\section{RHESSI Quiet Sun Data}

\subsection{Fan-beam Modulation Technique}\label{sec:method}

RHESSI makes images via a set of nine rotating modulation collimators RMCs,
whose resolution range logarithmically between 2.3$''$ and 183$''$
\citep{hurford2002}. Each of the grids also produces a coarser modulation,
depending on its thickness, on the order of the angular scale of the whole Sun.
To make use of this coarse modulation the spacecraft must point slightly away
from the Sun, the optimum effect occurring between 0.4$^\circ$ and 0.9$^\circ$
from disc centre. These operations interrupt the normal RHESSI program of flare
observations, so the quiet Sun mode is only used when solar activity is expected
to be at its lowest possible level. Data taken during these offpointing periods is
then fitted with the expected fan-beam sinusoidal modulation profile of a
uniform solar disc sized source \citep{hannah_fbm}, providing a measure of the
signal (or emission upper limit) above instrumental and terrestrial background.

\begin{figure}
\centering
\plotone{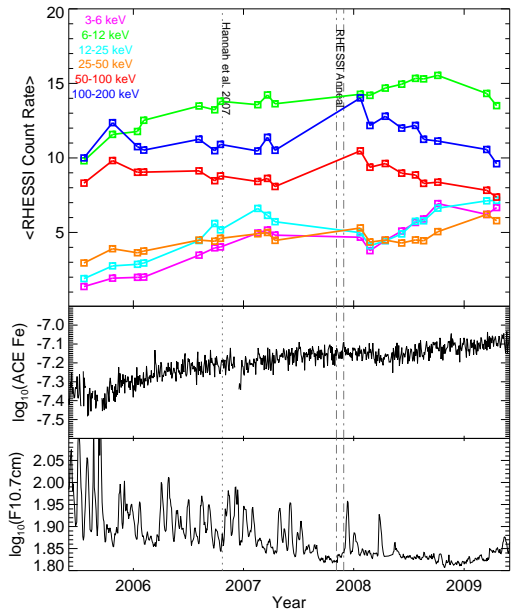}
\caption{
Time profile of the {\it RHESSI} count rate in different energy bands (top
panel)
averaged over detectors 1,3,4 and 6 and over the five minute intervals used to
determine the quiet Sun limits. The vertical line indicates the date up to which
the previous analysis had been done \citep{2007ApJ...659L..77H}. The dot-dash
lines indicate {\it RHESSI}'s first anneal (5-29 November 2007). The middle
panel
shows the Fe 270-450 MeV/nucleon rate for Galactic Cosmic Rays from
ACE/CRIS \citep{1998SSRv...86..285S}. The bottom panel shows the solar 10.7
cm radio flux, adjusted to 1AU (courtesy of the Canadian Space Weather
Forecast Centre).} \label{fig:vstime}
\end{figure}

In the present analysis we combine the older and newer data. The earlier data
consisted of seven intervals between 19~July 2005 and 23~October 2006, as
reported by \cite{2007ApJ...659L..77H}. The new data includes all of the RHESSI
quiet Sun observations following these, including the solar minimum between
Hale Cycles~23 and~24, and comprise an additional twelve periods from
12~February 2007 to 22~April 2009. The total number of observing sessions is
19, spanning 140 days.

For each of the offpointing periods we selected data with the criteria (i) GOES
SXR flux levels below the A1 level ($10^{-8}$ Wm$^{-2}$), (ii) no obvious GOES
or RHESSI time variations, and (iii) RHESSI background counting rates at the
minima in the latitude dependence due to cosmic radiation (see Figures 1 and 2
in \citet{2009ApJ...697...94M}). Each of the selected periods was split into
5~min intervals and then fitted with the expected fan-beam modulation profiles
\citep{hannah_fbm} for each detector and chosen energy band. This selection
resulted in 3,428 five-minutes intervals, a total of 11.9~days. We obtained a
fitted modulation amplitude for each interval, for each energy band, using the
subset of RMCs (numbers 1, 3, 4, and 6) best suited to this technique.

Figure~\ref{fig:vstime} summarises the data in the context of the background
cosmic-ray and solar variability. The mean rates are dominated by intrinsic
background sources, i.e. not by X-ray fluxes located within the imaging field of
view. During the entire interval of the RHESSI quiet Sun observations, the
galactic cosmic-ray flux was increasing towards record maximum levels, as
shown in the middle panel of the figure, based on Advanced Composition
Explorer (ACE) data \citep{1998SSRv...86..285S}. The increase of cosmic rays is
as expected from the solar-cycle modulation, and extends beyond the solar
activity minimum in late 2008, shown in the bottom panel by the solar 10.7~cm
radio flux, also as expected \citep{2009AGUFMSH13C..08M}. The low-energy
RHESSI analysis bands, excluding 6-12~keV, appear to show a similar upward
trend; this band contains a discrete instrumental spectral feature at about
10~keV. This keV wide feature is present in all detectors during both sunlight
and eclipse times and was speculated to be mostly due to the K-line emission
from radioactive decay in the germanium detectors \citep{2006ApJ...647.1480P}.
More specifically, via a private communication with A. Zoglauer and D. Smith, it
seems to be a due to cosmic protons causing electron capture decay producing
$^{71}$Ga fluorescence X-rays at about 10.4 keV.

\begin{deluxetable}{ccc}
\tablecolumns{3} \tablewidth{0pc} \tablecaption{\label{tab:results}The
weighted mean, and its associated standard deviation, of the {\it RHESSI} quiet
Sun photon flux. The previous values \citep{2007ApJ...659L..77H} are given in
brackets.} \tablehead{\colhead{Energy} &\colhead{Weighted Mean}
&\colhead{$\sigma$}\\ \colhead{keV} & \multicolumn{2}{c} {$\times 10^{-4}
\mathrm{ph}~\mathrm{s}^{-1}~\mathrm{cm}^{-2}~\mathrm{keV}^{-1}$}}
\startdata 3--6& -31.17 (330.99)&$\pm 170.19\; (\pm207.25)$ \\ 6--12& 5.97
(-5.24)&$\pm 4.75\; (\pm8.46)$ \\ 12--25& 0.51 (-0.73)&$\pm 0.94\; (\pm1.34)$
\\ 25--50& 0.02 (0.14)&$\pm 0.40\; (\pm0.63)$ \\ 50--100& -0.08 (-0.74)&$\pm
0.29\; (\pm0.54)$
\\ 100--200& -0.01 (-0.79)& $\pm 0.22\; (\pm0.42)$\\ \enddata
\end{deluxetable}

\begin{figure}
\centering
\plotone{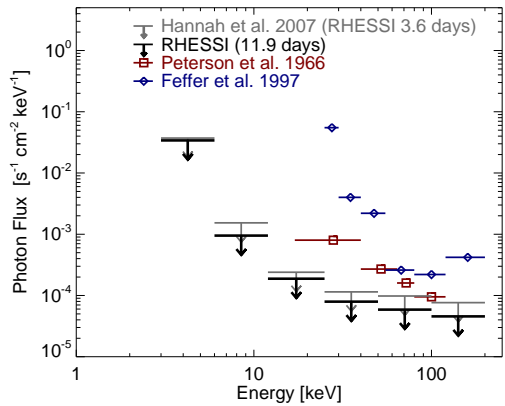}
\caption{
The RHESSI upper limits of the quiet Sun photon flux spectrum. The values are
the 2$\sigma$ limits, from the standard deviation of the weighted mean of the
four RMCs. The previous results using data during July 2005 to October 2006 is
also shown \citep{2007ApJ...659L..77H} as are the other HXR upper limits found
from \citet{peterson1966} and \citet{feffer1997}. }\label{fig:ph3200spec}
\end{figure}

At the higher energies (above about 50~keV) this cosmic-ray dependence
appears to decrease, and we speculate that the RHESSI background at these
energies is more closely associated with the trapped radiation around the Earth
than with the primary cosmic rays interacting in the Earth's atmosphere and
producing secondary radiations detectable at the RHESSI orbital altitude of
about 500~km. A further complication is the cumulative effect of radiation
damage to RHESSI's unshielded detectors over this period of low solar activity,
which increases background noise and reduces detector active volume. Due to
this no quiet Sun offpointing occurred in the second half of 2007, before a
detector anneal was conducted in November 2007 after which the detector
response recovered back to 2005 levels. Similarly, no quiet-Sun
offpointing was commanded after April 2009 due to the continued degradation
of RHESSI's detectors, despite the prolonged solar minimum. A second detector
anneal in March 2010 greatly improved the performance of the detectors,
returning it to early mission levels, but the Sun was no longer quiet.

For each energy band and detector the weighted mean and standard deviation
of the fitted amplitudes with their associated errors is calculated for all the time
intervals. These values are then converted from counts flux to photon flux using
the diagonal elements of RHESSI's detector response matrix. A final amplitude
and statistical error is then calculated, again using the weighted mean, from the
four values with errors per energy band. We find no significant signal in any
energy channel. Table~\ref{tab:results} gives the results in comparison with the
initial data of \cite{2007ApJ...659L..77H}. As expected, the further observations
has substantially reduced the derived limits, and the $>$1~$\sigma$ detection
found previously in the lowest (3-6~keV) band has become simply a limit.
Figure~\ref{fig:ph3200spec} shows these results graphically, in comparison with
the earlier results \citep{peterson1966,feffer1997}. These limits now become
the deepest limits for solar hard X-ray emission yet reported.

\section{Interpretations}

\subsection{Isothermal Emission}\label{sec:thermal}

The most natural interpretation of these observations would be as limits on
thermal sources in the corona, mainly free-free and free-bound continuum in the
HXR range. RHESSI also detects bound-bound emissions of Fe and Ni in
the 6-8~keV range \citep[e.g.][]{2004ApJ...605..921P}. Although the bulk of the
corona is too cool to produce thermal emission in the RHESSI range
above 3~keV, localised higher temperature emission (i.e. from bright points)
could easily provide emission in this energy range.

\begin{figure*}
\centering \plottwo{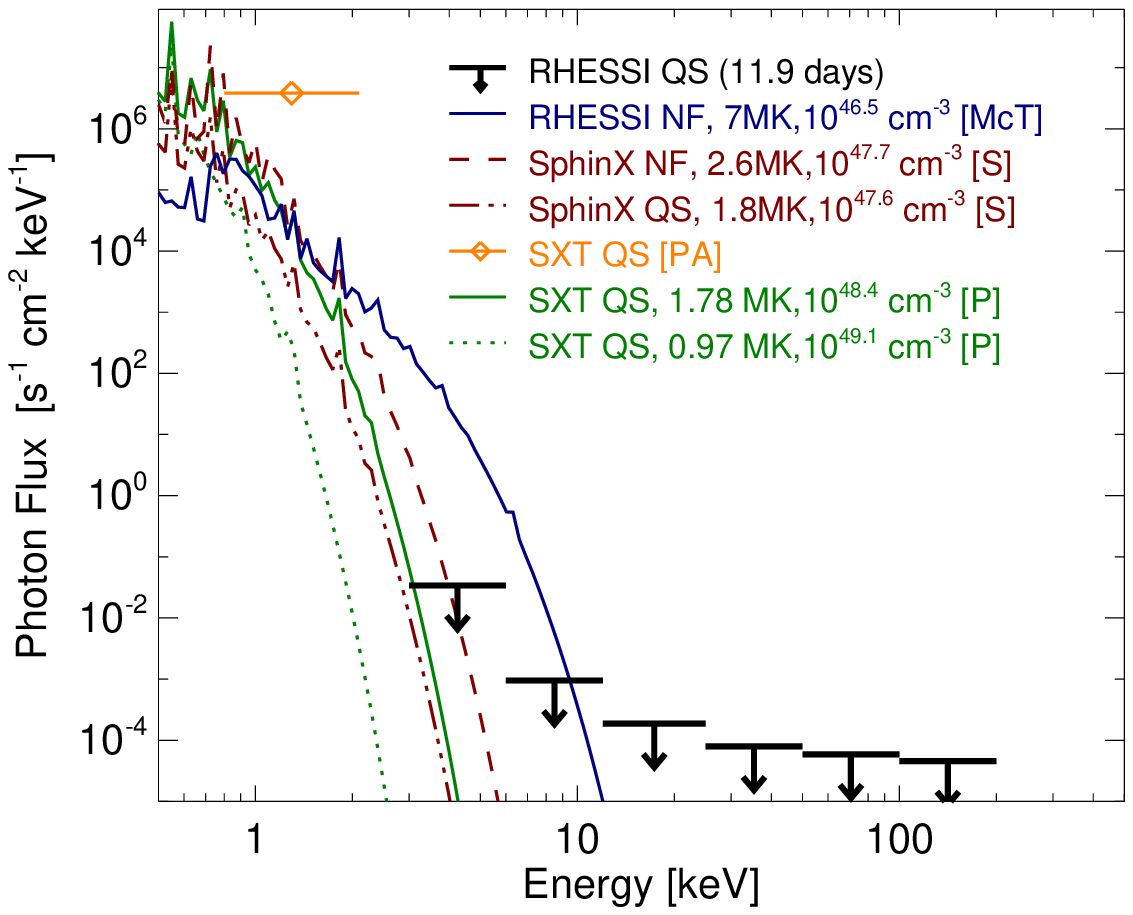}{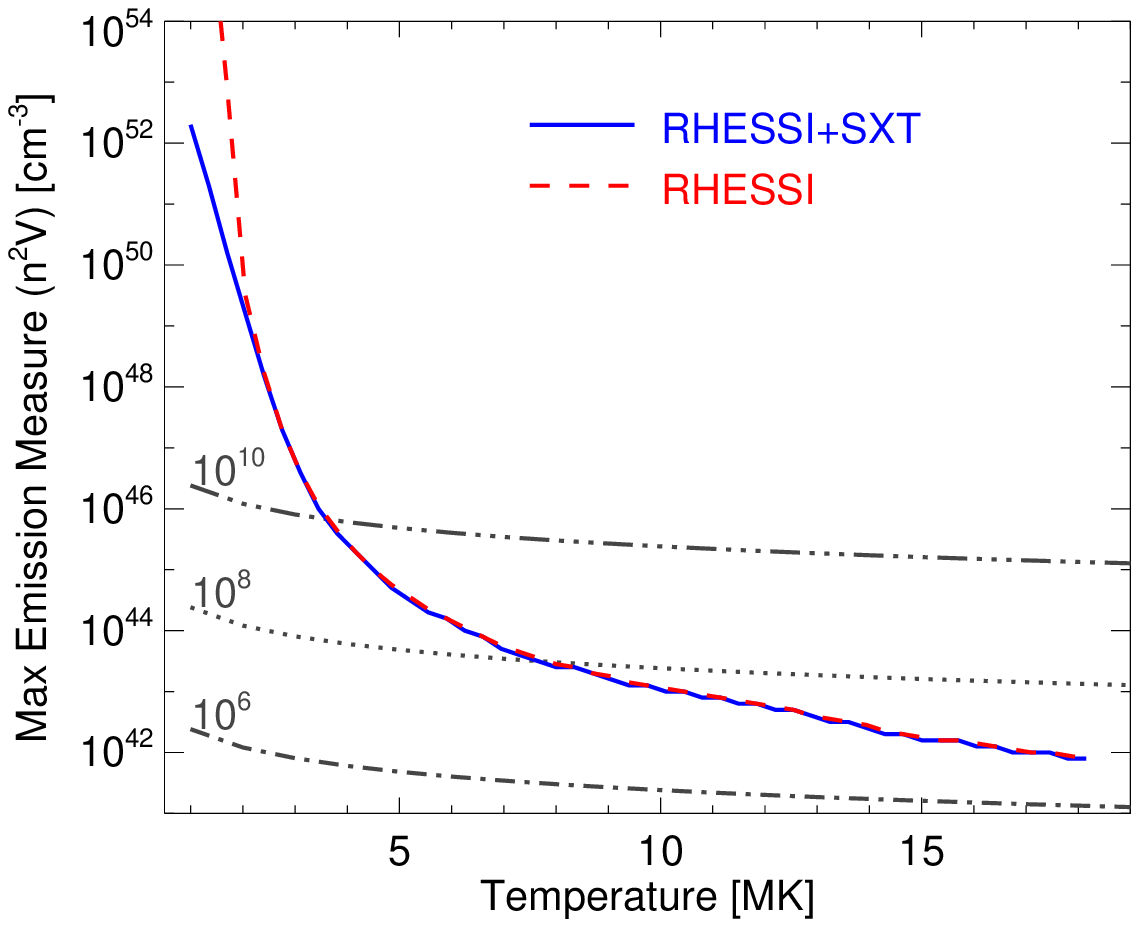} \caption{ (\emph{Left}) The {\it RHESSI} upper
limits compared to previously found thermal emission from the quiet Sun ([PA]
\citet{pevtsov2001}, [P] \citet{2000ApJ...528..537P}, [S]
\citet{2010EOSTr..91...73S}) and non-flaring active regions ([S]
\citet{2010EOSTr..91...73S}, [McT] \citet{2009ApJ...697...94M}). (\emph{Right})
The maximum emission measure as a function of temperature such that an
isothermal model produces a X-ray spectrum less than the {\it RHESSI} and {\it
Yohkoh/SXT} limits \citep{pevtsov2001}, shown in left panel. The area under the
curve is the possible parameter space consistent with the observations.
The dotted and dash-dotted grey lines indicate the emission measure
and temperature combination consistent with the coronal heating requirement
\citep{1977ARA&A..15..363W} with different background plasma densities
($n=10^{10},10^8,10^6$ cm$^{-3}$ from top to bottom).} \label{fig:emvst}
\end{figure*}

In the left panel of Figure~\ref{fig:emvst} we show the new RHESSI upper limits
in the context of previous quiet Sun and non-flaring active region observations.
Yohkoh/SXT produced a limiting value for the SXR quiet Sun \citep{pevtsov2001}
and this was used to find suitable isothermal model fits
\citep{2000ApJ...528..537P}. The SphinX observations of the end of Solar Cycle
23 have given preliminary estimates of a low, steady level of X-ray emission that
may provide the best characterisation of the background coronal emission
\citep{2010EOSTr..91...73S}. An isothermal fit was also made to this emission,
again shown in Figure~\ref{fig:emvst}. In both cases these quiet Sun isothermal
models are consistently lower than the RHESSI upper limits. Also shown are the
isothermal models fits during non-flaring quiescent active region times from
SphinX \citep{2010EOSTr..91...73S} and RHESSI \citep{2009ApJ...697...94M}. As
expected, the RHESSI upper limits are lower than the quiescent active region
emission.

In the right panel of Figure~\ref{fig:emvst} we have calculated the maximum
emission measure as a function of isothermal temperature which is consistent
with the RHESSI quiet Sun limits and the SXT constraint \citep{pevtsov2001}. We
find that this can be fitted with a polynomial of form

\begin{equation}
\log{EM}=52.97-15.25\log{T}+5.24\log^2{T}
\end{equation}

\noindent where $T$ is in units of MK and $EM$ is in units of cm$^{-3}$. Above
about 5 MK the emission measure is strongly constrained by the RHESSI upper
limits, with a maximum $<10^{44}$ cm$^{-3}$. For reference the standard
\cite{1988ApJ...325..442W} semi-empirical models of the solar wind, including
the corona, have emission measures in the range
0.8-9~$\times$~10$^{49}$~cm$^{-3}$, with peak temperatures in the range
1.42-1.64~MK and so are consistent with our limits. These however are
solar-wind models and therefore almost certainly underestimate both the
temperature and the emission measure of the steady-state quiet corona.

An additional constraint to the isothermal parameter space can be
obtained by considering the energy content being consistent with the coronal
heating requirement \citep{1977ARA&A..15..363W}. As a function of
temperature and for three assumed coronal densities ($n=10^{10},10^8,10^6$
cm$^{-3}$) we have estimated the emission measures, over plotting this in
Figure~\ref{fig:emvst}. The lowest density provides little further constraint but
the high densities suggest a maximum temperature of 7MK and about 4MK is
possible for densities of  $n=10^8$ cm$^{-3}$ and $n=10^{10}$ cm$^{-3}$
respectively.

\subsection{Nonthermal Thick-Target Emission}\label{sec:nonthermal}

The development of a solar flare involves nonthermal energy release, marked
for example by HXR and microwave emission, and the consequent increase of
coronal pressure in the flaring region. The pressure increase results from the
evaporation of chromospheric material to form the hot coronal plasma
responsible for SXR emission. The relationship between the nonthermal
component and the thermal component is well-understood observationally; the
peak SXR and HXR fluxes scale approximately linearly together within a factor of
10 or so over several decades \citep[e.g.][]{2001HvaOB..25...39V}. It is therefore
worthwhile to analyse our limits in terms of nonthermal bremsstrahlung,
especially since we do not know whether the flare relationship of nonthermal
and thermal processes holds for the quiet corona.

\begin{figure*}
\centering \plottwo{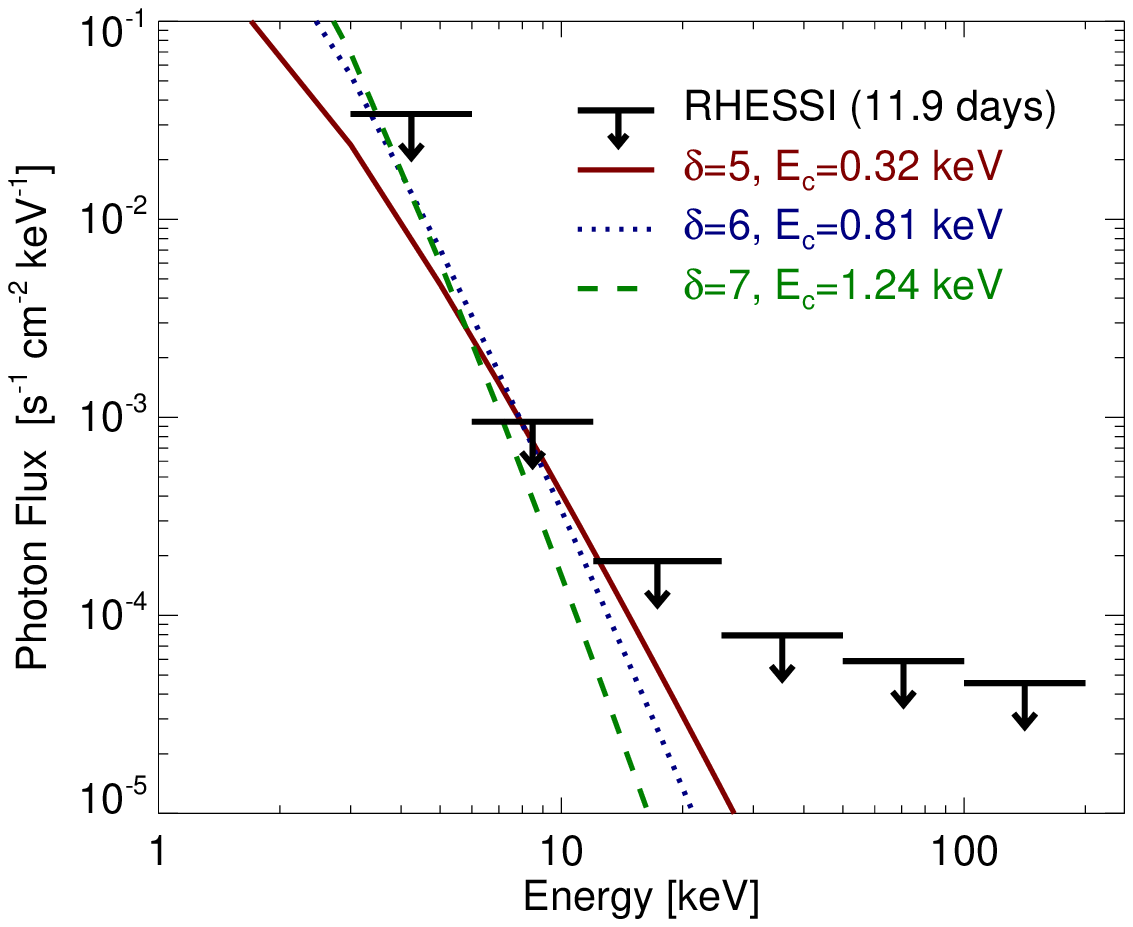}{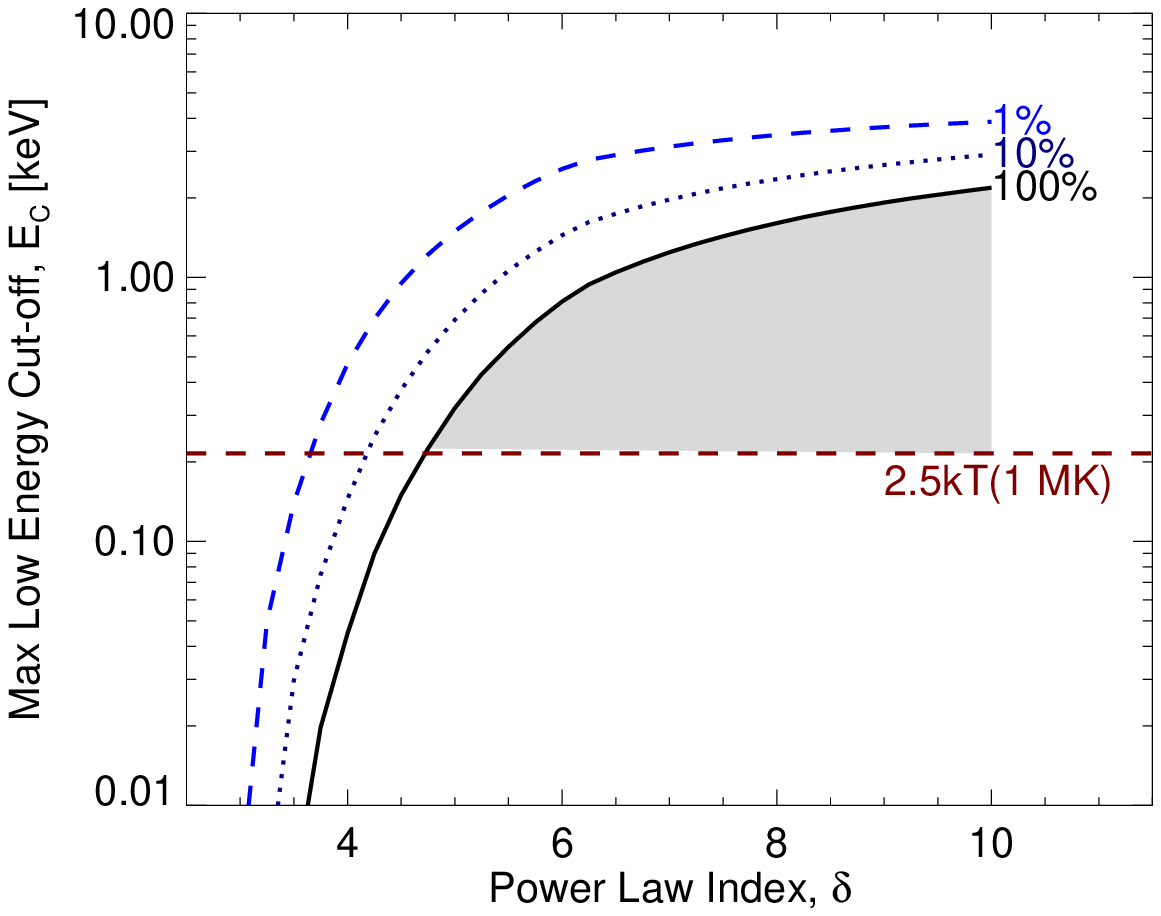} \caption{(\emph{Left}) The {\it RHESSI} upper
limits compared to thick-target model X-ray spectra from a power-law of
accelerated electrons of spectral index $\delta$ above cutoff energy $E_\mr{c}$
and consistent with the coronal heating requirement
\citep{1977ARA&A..15..363W}. (\emph{Right}) The possible nonthermal
parameters that could provide either 100\%, 10\% or 1\% of the coronal heating
requirement \citep{1977ARA&A..15..363W} while producing an X-ray spectrum
below the {\it RHESSI} limits (area below each curve). The horizontal line
indicates the possible lower limit to $E_\mr{C}$ based on a typical coronal
temperature \citep{2003ApJ...595L.119E}.}\label{fig:nncon}
\end{figure*}

We assume that there is a single power-law distribution of electrons
$f(E)\propto E^{-\delta}$ in the quiet Sun that produces HXR emission via
thick-target bremsstrahlung \citep{Brown1971}. Such a model is a good basis for
our limits since the corona contains mainly closed magnetic fields, and our long
integration times exceed the collisional loss times of electrons trapped within
them. This model has four parameters: the spectral index $\delta$, the energy
range over which the power-law extends (low energy cut off $E_\mathrm{C}$ to
maximum energy $E_\mathrm{M}$) and the total integrated electron flux,
$N=\int f(E)dE$ [electrons s$^{-1}$]. We fix the maximum energy at
$E_\mathrm{M}=1$ MeV as for the steep spectra and photon energy range we
are considering it has little effect. The remaining three parameters can be
further consolidated if we require a match to the assumed coronal heating
requirement $P_\mathrm{WN}=9\times10^{27}$ erg s$^{-1}$
\citep{1977ARA&A..15..363W}. The total integrated electron flux $N$ can then
be removed by rewriting it in terms of the power ($P=\int f(E) E dE$), i.e.

\begin{equation}
N=1.6\times10^{-9}\frac{P_\mathrm{WN}(\delta-2)}{E_\mathrm{C}(\delta-1)} \quad
\mathrm{electrons\; s}^{-1},
\end{equation}

\noindent where $E_\mathrm{C}$ is in keV. We can then investigate the
possible range of spectral index $\delta$ and low energy cutoff $E_\mathrm{C}$
that produce a thick-target bremsstrahlung spectrum $I(\epsilon)$ lower than
the RHESSI limits. Some example HXR spectra are shown in the left panel of
Figure~\ref{fig:nncon} which are consistent with the coronal heating
requirement and the RHESSI upper limits, using the numerical implementation of
\citet{2003ApJ...586..606H}. We can find the maximum possible low energy
cutoff that is possible for a range of spectral indices and this is shown in the
right panel of Figure~\ref{fig:nncon}. An additional parameter-space constraint
is of $E_\mr{C}=5kT/2$ as determined by the coronal thermal plasma
temperature $T$ \citep{2003ApJ...595L.119E}. With this we find that only steep
electron spectra ($\delta > 5$) are possible and that they extend down to very
low electron energies close to the thermal regime. Note that we have assumed
that the upper limits are solely due to nonthermal emission. An additional, and
highly likely, thermal component would reduce the nonthermal parameter-space
even further. We thus find a nanoflare coronal heating model based on flares
similar to nonthermal active region flares to be implausible.

\subsection{Thin-Target Emission}\label{sec:thin}
\begin{figure*}
\centering \plottwo{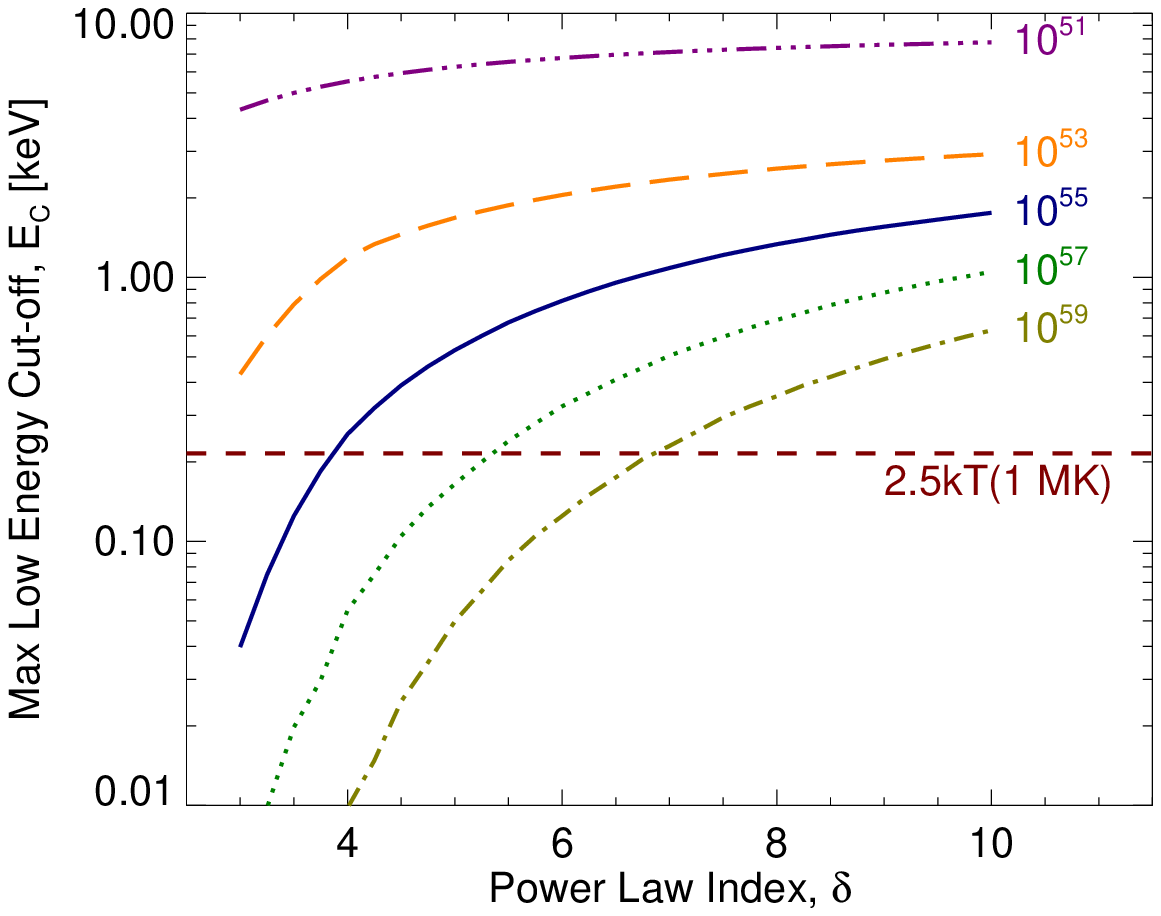}{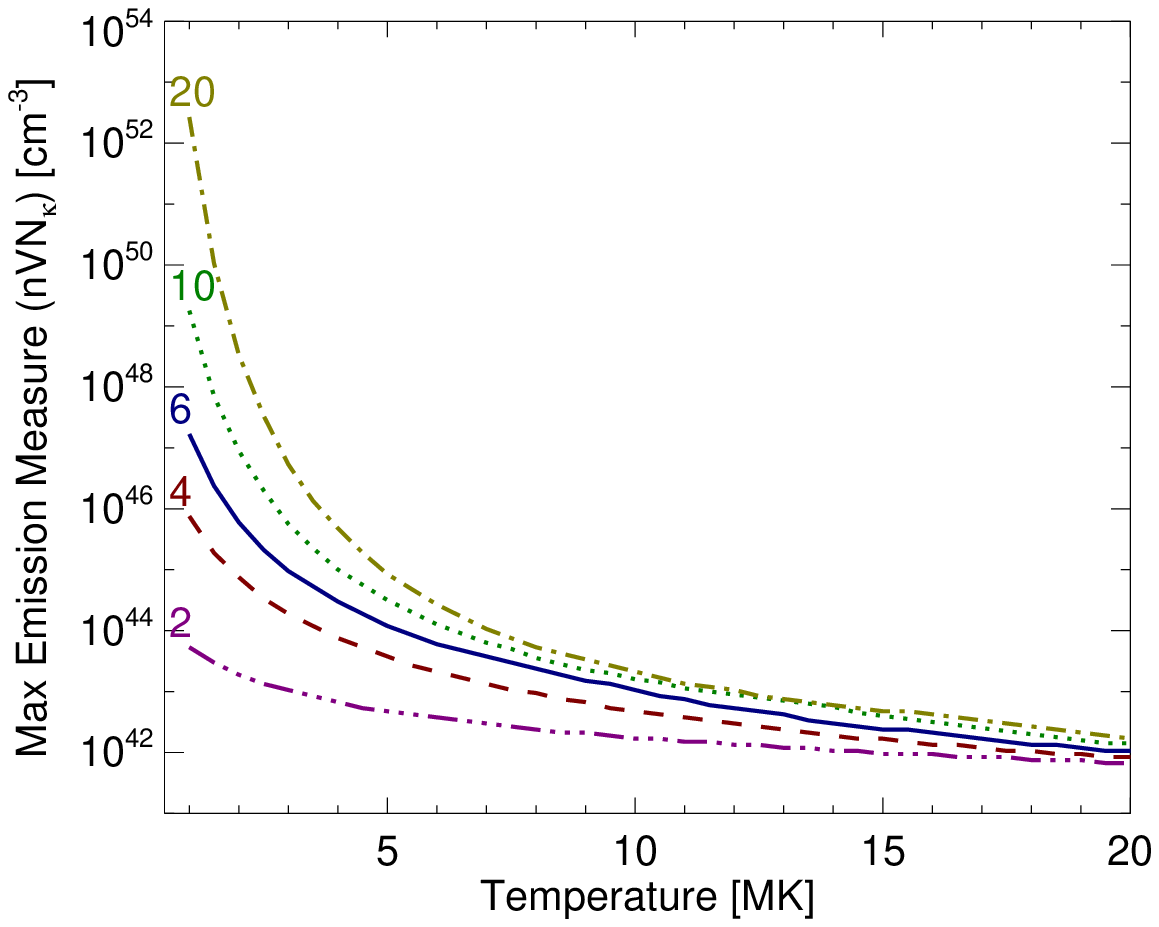} \caption{The possible parameter space for
(\emph{left}) a power-law distribution of accelerated electrons and
(\emph{right}) a kappa distribution that produce thin-target X-ray emission
consistent with the RHESSI upper limits. The different lines denote the maximum
possible in the parameter space for different values of the (\emph{left})
normalisation parameter $nVN$ and (\emph{right}) kappa parameter $\kappa$.
(\emph{Left}) The horizontal line indicates the possible lower limit to
$E_\mr{C}$ based on a typical coronal temperature
\citep{2003ApJ...595L.119E}.}\label{fig:thin}
\end{figure*}

Another likely emission mechanism to produce quiet Sun HXRs is
via a coronal thin-target process \citep[e.g.][]{1976SoPh...50..153L}, where
energised
electrons would continuously emit via bremsstrahlung interactions with the
coronal plasma but would lose little energy doing so (unlike the complete
energy loss through collisions with the denser chromosphere in the thick-target
case \S\ref{sec:nonthermal}). For these models we cannot use the coronal
heating requirement to constrain the parameter space as there is no substantial
energy loss to heat the background plasma. We consider two models both of
which are functions of three parameters. We again consider a power law
distribution of electrons with spectral index $\delta$ above a low energy cut-off
$E_\mathrm{C}$ (extending up to energy of 1 MeV), this time normalised by the
product
of the plasma density, volume of emitting plasma and integrated electrons flux
($nVN$ [cm$^{-2}$ s$^{-1}$]). The parameter space ($E_\mathrm{C}$, $\delta$)
of this model, for different values of the normalisation, that produces thin-target
emission $I(\epsilon)$ less than the RHESSI upper limits are shown in Figure
\ref{fig:thin}. As the normalisation factor increases the maximum low energy
cut-off sharply decreases, requiring $\delta >7$ for $nVN=10^{59}$
cm$^{-2}$s$^{-1}$ again once the additional constraint of
\citet{2003ApJ...595L.119E} is included.

The second model we consider is a kappa distribution which can fit
observed in-situ solar wind distribution and some coronal flare spectra
\citep{2009A&A...497L..13K}. The distribution is a function of emission measure
$nVN_\kappa$ (not the same as the isothermal emission measure $n^2V$ as
$N_\kappa$, the electron density in the kappa distribution, is included with the
background plasma density $n$) temperature $T$ and the kappa parameter
$\kappa$, numerically implemented using the version from
\citet{2009A&A...497L..13K}. The kappa parameter adds a high-energy tail to the
thermal Maxwellian, approaching a powerlaw at high energies for low $\kappa$.
The emission measure temperature parameter space for various values of
$\kappa$ that produce thin-target emission $I(\epsilon)$ consistent with the
RHESSI limits are shown in the right panel of Figure \ref{fig:thin}. Low values of
$\kappa<6$ (a flat tail) greatly reduce the possible emission measure since we
have more high energy electrons in the tail to produce X-rays. For larger values
$\kappa \ge 20$ we approach the isothermal constraints shown in
Figure~\ref{fig:emvst}.

\subsection{Axions}\label{sec:other}

The flux of axions thought to be produced in the Sun's core have a  mean energy
of 4.2~keV in a roughly blackbody distribution
\citep{1989PhRvD..39.2089V,2007JCAP...04..010A} and convert directly to
photons of the same energy with probability proportional to $(\int{B_\perp
dl})^2$ (the perpendicular magnetic field encountered) and
$g_{a\gamma\gamma}^2$, an unknown coupling constant. The unique
parameter space available to the RHESSI limits further constrain this coupling.
The limits in 3-6 keV presented in this paper are about 20\% smaller than those
from the previous analysis \citep{2007ApJ...659L..77H}. Assuming that these
limits are exclusively due to axions then we find our limits to be lower than the
X-ray emission predicted for light axion conversion in a simple dipole field with
$g_\mr{a\gamma\gamma}=10^{-10}~\mr{GeV}^{-1}$ \citep{carlson1996}. A
smaller $g_\mr{a\gamma\gamma}$ or a modified magnetic field model could
produce X-ray emission within our limits. For the scenario of massive
Kaluza-Klein axions our new X-ray upper limits still produce $g_\mr{a \gamma
\gamma} \ll 6 \times 10^{-15}~\mr{GeV}^{-1}$, using the method of
\citet{zioutas2004}, since the X-ray luminosity is proportional to $g_\mr{a
\gamma \gamma}^4$.

A better treatment of this problem would require more complete knowledge of
the perpendicular magnetic fields encountered by the axions fleeing the Sun.
This field would not be expected to vary during solar minimum except for
statistical fluctuations of the magnetic field in the quiet Sun. In the presence of
higher levels of activity, and stronger localised magnetic fields, strong spatial
and temporal variations would become evident. To the extent that the axion
spectral signature cannot be disentangled, the normal mechanisms of solar
magnetic activity could easily outweigh the axion source intensity.

\section{Conclusions}\label{sec:discuss}

The RHESSI observations reported here give the best upper limits yet on solar
X-ray emission, at the quietest times, above 3~keV. These limits constrain
models of coronal heating that require high temperatures or nonthermal
particles and possible coronal thin-target emission. In all instances this
was considered in terms of a spatially and temporally averaged emission, a
``typical'' nanoflare, whereas a distribution of nanoflares could easily produce
individual events brighter than the RHESSI upper limits for short periods of time.
For the high-temperature tail of a DEM consistent with nanoflare heating, we
find that the fraction of emission measure above 5~MK must be $\lesssim
10^{-6}$ of the peak of the DEM needed for the quiet corona, crudely estimated
at 2~$\times$~10$^{50}$~cm$^{-3}$ for a coronal base density of
10$^{9}$~cm$^{-3}$. Nanoflare models \citep[e.g.][]{2008ApJ...682.1351K}
involve many interrelated free parameters at present, and we hope that our
strong limits will be incorporated into future theoretical work. Further
parameter-space constraints result if we interpret our limits in terms of
nonthermal
bremsstrahlung from accelerated electrons. Here the limits force the spectral
index $\delta$ to be steeper than about~5 for any physically meaningful
low-energy cutoff energy $E_\mathrm{c}$. With this nonthermal interpretation,
heating via particle acceleration, we demonstrated that it was unlikely that
nanoflares could heat the corona in a manner akin to heating in ordinary flares.

The RHESSI solar observations we report here, though the best ever achieved in
the HXR range, could be greatly improved since RHESSI (and most other solar
instruments) are not optimised for faint sources. One approach would be using
focusing optics, allowing quiet regions of the corona to be isolated with high
sensitivity and wide dynamic range, possible with technology such as
\emph{FOXSI} \citep{2009SPIE.7437E...4K} sounding rocket and NuSTAR
\citep{2010HEAD...11.4601H} satellite instruments, both scheduled for launch.
Such observations would not only allow us to investigate the existence
and nature of a quiet Sun accelerated electron population but would greatly
benefit our understanding of energy release and transport processes in
active-region flares.

\section{Acknowledgements}

IGH is supported by a STFC rolling grant and by the European Commission
through the SOLAIRE Network (MTRN-CT-2006-035484). This work was
supported in part by NASA contract NAS5-98033. R. Lin was also supported in
part by the WCU grant (No. R31-10016) funded by the Korean Ministry of
Education, Science and Technology.

% \bibliographystyle{apj}
% \bibliography{refs}

\begin{thebibliography}{38}
\expandafter\ifx\csname natexlab\endcsname\relax\def\natexlab#1{#1}\fi

\bibitem[{{Andriamonje} {et~al.}(2007){Andriamonje}, {60 co-authors}, \& the
  (CAST~Collaboration)}]{2007JCAP...04..010A}
{Andriamonje}, S., {60 co-authors}, \& the (CAST~Collaboration). 2007, Journal
  of Cosmology and Astro-Particle Physics, 4, 10

\bibitem[{{Brown}(1971)}]{Brown1971}
{Brown}, J.~C. 1971, \solphys, 18, 489

\bibitem[{{Cargill}(1994)}]{1994ApJ...422..381C}
{Cargill}, P.~J. 1994, \apj, 422, 381

\bibitem[{{Cargill} \& {Klimchuk}(1997)}]{1997ApJ...478..799C}
{Cargill}, P.~J., \& {Klimchuk}, J.~A. 1997, \apj, 478, 799

\bibitem[{{Carlson} \& {Tseng}(1996)}]{carlson1996}
{Carlson}, E.~D., \& {Tseng}, L.~S. 1996, Physics Letters B, 365, 193

\bibitem[{{Churazov} {et~al.}(2008){Churazov}, {Sazonov}, {Sunyaev}, \&
  {Revnivtsev}}]{2008MNRAS.385..719C}
{Churazov}, E., {Sazonov}, S., {Sunyaev}, R., \& {Revnivtsev}, M. 2008, \mnras,
  385, 719

\bibitem[{{Emslie}(2003)}]{2003ApJ...595L.119E}
{Emslie}, A.~G. 2003, \apjl, 595, L119

\bibitem[{{Feffer} {et~al.}(1997){Feffer}, {Lin}, {Slassi-Sennou}, {McBride},
  {Primbsch}, {Zimmer}, {Pelling}, {Pehl}, {Madden}, {Malone}, {Cork}, {Luke},
  {Vedrenne}, \& {Cotin}}]{feffer1997}
{Feffer}, P.~T., {Lin}, R.~P., {Slassi-Sennou}, S., {McBride}, S., {Primbsch},
  J.~H., {Zimmer}, G., {Pelling}, R.~M., {Pehl}, R., {Madden}, N., {Malone},
  D., {Cork}, C., {Luke}, P., {Vedrenne}, G., \& {Cotin}, F. 1997, \solphys,
  171, 419

\bibitem[{{Hannah} {et~al.}(2008){Hannah}, {Christe}, {Krucker}, {Hurford},
  {Hudson}, \& {Lin}}]{2008ApJ...677..704H}
{Hannah}, I.~G., {Christe}, S., {Krucker}, S., {Hurford}, G.~J., {Hudson},
  H.~S., \& {Lin}, R.~P. 2008, \apj, 677, 704

\bibitem[{{Hannah} {et~al.}(2007{\natexlab{a}}){Hannah}, {Hurford}, {Hudson},
  \& {Lin}}]{hannah_fbm}
{Hannah}, I.~G., {Hurford}, G.~J., {Hudson}, H.~S., \& {Lin}, R.~P.
  2007{\natexlab{a}}, Review of Scientific Instruments, 78, 10

\bibitem[{{Hannah} {et~al.}(2007{\natexlab{b}}){Hannah}, {Hurford}, {Hudson},
  {Lin}, \& {van Bibber}}]{2007ApJ...659L..77H}
{Hannah}, I.~G., {Hurford}, G.~J., {Hudson}, H.~S., {Lin}, R.~P., \& {van
  Bibber}, K. 2007{\natexlab{b}}, \apjl, 659, L77

\bibitem[{{Harrison} {et~al.}(2010){Harrison}, {Boggs}, {Christensen}, {Craig},
  {Hailey}, {Stern}, {Zhang}, \& {NuSTAR Science Team}}]{2010HEAD...11.4601H}
{Harrison}, F., {Boggs}, S., {Christensen}, F., {Craig}, W., {Hailey}, C.,
  {Stern}, D., {Zhang}, W., \& {NuSTAR Science Team}. 2010, in Bulletin of the
  American Astronomical Society, Vol.~41, Bulletin of the American Astronomical
  Society, 737--+

\bibitem[{{Holman}(2003)}]{2003ApJ...586..606H}
{Holman}, G.~D. 2003, \apj, 586, 606

\bibitem[{{Hurford} {et~al.}(2002){Hurford}, {Schmahl}, {Schwartz}, {Conway},
  {Aschwanden}, {Csillaghy}, {Dennis}, {Johns-Krull}, {Krucker}, {Lin},
  {McTiernan}, {Metcalf}, {Sato}, \& {Smith}}]{hurford2002}
{Hurford}, G.~J., {Schmahl}, E.~J., {Schwartz}, R.~A., {Conway}, A.~J.,
  {Aschwanden}, M.~J., {Csillaghy}, A., {Dennis}, B.~R., {Johns-Krull}, C.,
  {Krucker}, S., {Lin}, R.~P., {McTiernan}, J., {Metcalf}, T.~R., {Sato}, J.,
  \& {Smith}, D.~M. 2002, \solphys, 210, 61

\bibitem[{{Ka{\v s}parov{\'a}} \& {Karlick{\'y}}(2009)}]{2009A&A...497L..13K}
{Ka{\v s}parov{\'a}}, J., \& {Karlick{\'y}}, M. 2009, \aap, 497, L13

\bibitem[{{Klimchuk} {et~al.}(2008){Klimchuk}, {Patsourakos}, \&
  {Cargill}}]{2008ApJ...682.1351K}
{Klimchuk}, J.~A., {Patsourakos}, S., \& {Cargill}, P.~J. 2008, \apj, 682, 1351

\bibitem[{{Krucker} {et~al.}(2009){Krucker}, {Christe}, {Glesener}, {McBride},
  {Turin}, {Glaser}, {Saint-Hilaire}, {Delory}, {Lin}, {Gubarev}, {Ramsey},
  {Terada}, {Ishikawa}, {Kokubun}, {Saito}, {Takahashi}, {Watanabe},
  {Nakazawa}, {Tajima}, {Masuda}, {Minoshima}, \&
  {Shomojo}}]{2009SPIE.7437E...4K}
{Krucker}, S., {Christe}, S., {Glesener}, L., {McBride}, S., {Turin}, P.,
  {Glaser}, D., {Saint-Hilaire}, P., {Delory}, G., {Lin}, R.~P., {Gubarev}, M.,
  {Ramsey}, B., {Terada}, Y., {Ishikawa}, S., {Kokubun}, M., {Saito}, S.,
  {Takahashi}, T., {Watanabe}, S., {Nakazawa}, K., {Tajima}, H., {Masuda}, S.,
  {Minoshima}, T., \& {Shomojo}, M. 2009, in Society of Photo-Optical
  Instrumentation Engineers (SPIE) Conference Series, Vol. 7437, Society of
  Photo-Optical Instrumentation Engineers (SPIE) Conference Series

\bibitem[{{Lin} {et~al.}(2002){Lin}, {Dennis}, {Hurford}, {Smith}, {Zehnder},
  {Harvey}, {Curtis}, {Pankow}, {Turin}, {Bester}, {Csillaghy}, {Lewis},
  {Madden}, {van Beek}, {Appleby}, {Raudorf}, {McTiernan}, {Ramaty}, {Schmahl},
  {Schwartz}, {Krucker}, {Abiad}, {Quinn}, {Berg}, {Hashii}, {Sterling},
  {Jackson}, {Pratt}, {Campbell}, {Malone}, {Landis}, {Barrington-Leigh},
  {Slassi-Sennou}, {Cork}, {Clark}, {Amato}, {Orwig}, {Boyle}, {Banks},
  {Shirey}, {Tolbert}, {Zarro}, {Snow}, {Thomsen}, {Henneck}, {McHedlishvili},
  {Ming}, {Fivian}, {Jordan}, {Wanner}, {Crubb}, {Preble}, {Matranga}, {Benz},
  {Hudson}, {Canfield}, {Holman}, {Crannell}, {Kosugi}, {Emslie}, {Vilmer},
  {Brown}, {Johns-Krull}, {Aschwanden}, {Metcalf}, \& {Conway}}]{lin2002}
{Lin}, R.~P., {Dennis}, B.~R., {Hurford}, G.~J., {Smith}, D.~M., {Zehnder}, A.,
  {Harvey}, P.~R., {Curtis}, D.~W., {Pankow}, D., {Turin}, P., {Bester}, M.,
  {Csillaghy}, A., {Lewis}, M., {Madden}, N., {van Beek}, H.~F., {Appleby}, M.,
  {Raudorf}, T., {McTiernan}, J., {Ramaty}, R., {Schmahl}, E., {Schwartz}, R.,
  {Krucker}, S., {Abiad}, R., {Quinn}, T., {Berg}, P., {Hashii}, M.,
  {Sterling}, R., {Jackson}, R., {Pratt}, R., {Campbell}, R.~D., {Malone}, D.,
  {Landis}, D., {Barrington-Leigh}, C.~P., {Slassi-Sennou}, S., {Cork}, C.,
  {Clark}, D., {Amato}, D., {Orwig}, L., {Boyle}, R., {Banks}, I.~S., {Shirey},
  K., {Tolbert}, A.~K., {Zarro}, D., {Snow}, F., {Thomsen}, K., {Henneck}, R.,
  {McHedlishvili}, A., {Ming}, P., {Fivian}, M., {Jordan}, J., {Wanner}, R.,
  {Crubb}, J., {Preble}, J., {Matranga}, M., {Benz}, A., {Hudson}, H.,
  {Canfield}, R.~C., {Holman}, G.~D., {Crannell}, C., {Kosugi}, T., {Emslie},
  A.~G., {Vilmer}, N., {Brown}, J.~C., {Johns-Krull}, C., {Aschwanden}, M.,
  {Metcalf}, T., \& {Conway}, A. 2002, \solphys, 210, 3

\bibitem[{{Lin} \& {Hudson}(1976)}]{1976SoPh...50..153L}
{Lin}, R.~P., \& {Hudson}, H.~S. 1976, \solphys, 50, 153

\bibitem[{{MacKinnon}(2007)}]{2007A&A...462..763M}
{MacKinnon}, A.~L. 2007, \aap, 462, 763

\bibitem[{{McTiernan}(2009)}]{2009ApJ...697...94M}
{McTiernan}, J.~M. 2009, \apj, 697, 94

\bibitem[{{Mewaldt} {et~al.}(2009){Mewaldt}, {Davis}, {Lave}, {Leske},
  {Wiedenbeck}, {Binns}, {Christian}, {Cummings}, {de Nolfo}, {Israel},
  {Stone}, \& {von Rosenvinge}}]{2009AGUFMSH13C..08M}
{Mewaldt}, R.~A., {Davis}, A.~J., {Lave}, K.~A., {Leske}, R.~A., {Wiedenbeck},
  M.~E., {Binns}, W.~R., {Christian}, E.~R., {Cummings}, A.~C., {de Nolfo},
  G.~A., {Israel}, M.~H., {Stone}, E.~C., \& {von Rosenvinge}, T.~T. 2009, AGU
  Fall Meeting Abstracts, C8+

\bibitem[{{Orlando} {et~al.}(2009){Orlando}, {Giglietto}, \& {for the Fermi
  Large Area Telescope Collaboration}}]{2009arXiv0912.3775O}
{Orlando}, E., {Giglietto}, N., \& {for the Fermi Large Area Telescope
  Collaboration}. 2009, ArXiv e-prints

\bibitem[{{Parker}(1988)}]{parker1988}
{Parker}, E.~N. 1988, \apj, 330, 474

\bibitem[{{Peres} {et~al.}(2000){Peres}, {Orlando}, {Reale}, {Rosner}, \&
  {Hudson}}]{2000ApJ...528..537P}
{Peres}, G., {Orlando}, S., {Reale}, F., {Rosner}, R., \& {Hudson}, H. 2000,
  \apj, 528, 537

\bibitem[{{Peterson} {et~al.}(1966){Peterson}, {Schwartz}, {Pelling}, \&
  {McKenzie}}]{peterson1966}
{Peterson}, L.~E., {Schwartz}, D.~A., {Pelling}, R.~M., \& {McKenzie}, D. 1966,
  \jgr, 71, 5778

\bibitem[{{Pevtsov} \& {Acton}(2001)}]{pevtsov2001}
{Pevtsov}, A.~A., \& {Acton}, L.~W. 2001, \apj, 554, 416

\bibitem[{{Phillips}(2004)}]{2004ApJ...605..921P}
{Phillips}, K.~J.~H. 2004, \apj, 605, 921

\bibitem[{{Phillips} {et~al.}(2006){Phillips}, {Chifor}, \&
  {Dennis}}]{2006ApJ...647.1480P}
{Phillips}, K.~J.~H., {Chifor}, C., \& {Dennis}, B.~R. 2006, \apj, 647, 1480

\bibitem[{{Seckel} {et~al.}(1991){Seckel}, {Stanev}, \& {Gaisser}}]{seckel1991}
{Seckel}, D., {Stanev}, T., \& {Gaisser}, T.~K. 1991, \apj, 382, 652

\bibitem[{{Sikivie}(1983)}]{sikivie1983}
{Sikivie}, P. 1983, Physical Review Letters, 51, 1415

\bibitem[{{Stone} {et~al.}(1998){Stone}, {Cohen}, {Cook}, {Cummings}, {Gauld},
  {Kecman}, {Leske}, {Mewaldt}, {Thayer}, {Dougherty}, {Grumm}, {Milliken},
  {Radocinski}, {Wiedenbeck}, {Christian}, {Shuman}, {Trexel}, {von
  Rosenvinge}, {Binns}, {Crary}, {Dowkontt}, {Epstein}, {Hink}, {Klarmann},
  {Lijowski}, \& {Olevitch}}]{1998SSRv...86..285S}
{Stone}, E.~C., {Cohen}, C.~M.~S., {Cook}, W.~R., {Cummings}, A.~C., {Gauld},
  B., {Kecman}, B., {Leske}, R.~A., {Mewaldt}, R.~A., {Thayer}, M.~R.,
  {Dougherty}, B.~L., {Grumm}, R.~L., {Milliken}, B.~D., {Radocinski}, R.~G.,
  {Wiedenbeck}, M.~E., {Christian}, E.~R., {Shuman}, S., {Trexel}, H., {von
  Rosenvinge}, T.~T., {Binns}, W.~R., {Crary}, D.~J., {Dowkontt}, P.,
  {Epstein}, J., {Hink}, P.~L., {Klarmann}, J., {Lijowski}, M., \& {Olevitch},
  M.~A. 1998, Space Science Reviews, 86, 285

\bibitem[{{Sylwester} {et~al.}(2010){Sylwester}, {Kowalinski}, {Gburek},
  {Siarkowski}, {Kuzin}, {Farnik}, {Reale}, \&
  {Phillips}}]{2010EOSTr..91...73S}
{Sylwester}, J., {Kowalinski}, M., {Gburek}, S., {Siarkowski}, M., {Kuzin}, S.,
  {Farnik}, F., {Reale}, F., \& {Phillips}, K.~J.~H. 2010, EOS Transactions,
  91, 73

\bibitem[{{van Bibber} {et~al.}(1989){van Bibber}, {McIntyre}, {Morris}, \&
  {Raffelt}}]{1989PhRvD..39.2089V}
{van Bibber}, K., {McIntyre}, P.~M., {Morris}, D.~E., \& {Raffelt}, G.~G. 1989,
  \prd, 39, 2089

\bibitem[{{Veronig} {et~al.}(2001){Veronig}, {Vr{\v s}nak}, {Temmer},
  {Magdaleni{\'c}}, \& {Hanslmeier}}]{2001HvaOB..25...39V}
{Veronig}, A., {Vr{\v s}nak}, B., {Temmer}, M., {Magdaleni{\'c}}, J., \&
  {Hanslmeier}, A. 2001, Hvar Observatory Bulletin, 25, 39

\bibitem[{{Withbroe}(1988)}]{1988ApJ...325..442W}
{Withbroe}, G.~L. 1988, \apj, 325, 442

\bibitem[{{Withbroe} \& {Noyes}(1977)}]{1977ARA&A..15..363W}
{Withbroe}, G.~L., \& {Noyes}, R.~W. 1977, \araa, 15, 363

\bibitem[{{Zioutas} {et~al.}(2004){Zioutas}, {Dennerl}, {DiLella}, {Hoffmann},
  {Jacoby}, \& {Papaevangelou}}]{zioutas2004}
{Zioutas}, K., {Dennerl}, K., {DiLella}, L., {Hoffmann}, D.~H.~H., {Jacoby},
  J., \& {Papaevangelou}, T. 2004, \apj, 607, 575

\end{thebibliography}

\end{document}